# Crossing-Free Acyclic Hamiltonian Path Completion for Planar $st$-Digraphs

Tamara Mchedlidze, Antonios Symvonis

Dept. of Mathematics, National Technical University of Athens, Athens, Greece.
{mchet,symvonis}@math.ntua.gr

**Abstract.** In this paper we study the problem of existence of a crossing-free acyclic hamiltonian path completion (for short, HP-completion) set for embedded upward planar digraphs. In the context of book embeddings, this question becomes: given an embedded upward planar digraph $G$, determine whether there exists an upward 2-page book embedding of $G$ preserving the given planar embedding. Given an embedded $st$-digraph $G$ which has a crossing-free HP-completion set, we show that there always exists a crossing-free HP-completion set with at most two edges per face of $G$. For an embedded $N$-free upward planar digraph $G$, we show that there always exists a crossing-free acyclic HP-completion set for $G$ which, moreover, can be computed in linear time. For a width-$k$ embedded planar $st$-digraph $G$, we show that we can be efficiently test whether $G$ admits a crossing-free acyclic HP-completion set.

## 1 Introduction

A *k-page book* is a structure consisting of a line, referred to as *spine*, and of $k$ half-planes, referred to as *pages*, that have the spine as their common boundary. A *book embedding* of a graph $G$ is a drawing of $G$ on a book such that the vertices are aligned along the spine, each edge is entirely drawn on a single page, and edges do not cross each other. If we are interested only in two-dimensional structures we have to concentrate on 2-page book embeddings and to allow spine crossings. These embeddings are also referred to as 2-page *topological* book embeddings.

For acyclic digraphs, an upward book embedding can be considered to be a book embedding in which the spine is vertical and all edges are drawn monotonically increasing in the upward direction. As a consequence, in an upward book embedding of an acyclic digraph $G$ the vertices of $G$ appear along the spine in topological order. If $G$ is planar upward digraph and an upward embedding of $G$ on the plane is given, we are interested to determine a 2-page upward topological book embedding of $G$ which preserves its plane embedding and has minimum number of spine crossings. Giordano et al. [10] showed that an embedded upward planar digraph always admits an upward topological 2-page book embedding (which preserves its plane embedding) with at most one spine crossing per edge. However, in their work no effort was made to minimize the total number of spine crossings.

The *acyclic hamiltonian path completion with crossing minimization problem* (*Acyclic-HPCCM*) was inspired by its equivalence with the problem of determining an upward 2-page topological book embedding with a minimum number of spine crossings for an embedded planar $st$-digraph [18].

In the *hamiltonian path completion problem* (*HPC*) we are given a graph[1] $G$ and we are asked to identify a set of edges $S$ (refereed to as an *HP-completion set*) such that, when the edges of $S$ are embedded on $G$ they turn it to a hamiltonian graph, that is, a graph containing a hamiltonian path[2]. The resulting hamiltonian graph $G_S$ is referred to as the *HP-completed graph* of $G$. When we treat the HP-completion problem as an optimization problem, we are interested in HP-completion sets of minimum size.

When the input graph $G$ is an embedded planar digraph, an HP-completion set $S$ for $G$ must be naturally extended to include an embedding of its edges on the plane, yielding to an embedded HP-completed digraph $G_S$. In general, $G_S$ is not planar, and thus, it is natural to attempt to minimize the number of edge crossings of the embedding of the HP-completed digraph $G_S$ instead of the size of the HP-completion set $S$. This problem is known as *HP-completion with crossing minimization problem* (*HPCCM*) and was first defined in [18].

---

[1] In this paper, we assume that $G$ is directed.
[2] In the literature, a *hamiltonian graph* is traditionally referred to as a graph containing a hamiltonian cycle. In this paper, we refer to a hamiltonian graph as a graph containing a hamiltonian path.

When the input digraph $G$ is acyclic, we can insist on HP-completion sets which leave the HP-completed digraph $G'$ also acyclic. We refer to this version of the problem as the *Acyclic-HPC problem*. Analogously, we define the *acyclic-HPCCM* which, as stated above, is equivalent to determining 2-page upward topological book embeddings with minimum number of spine crossings for embedded upward planar digraphs.

When dealing with the acyclic-HPCCM problem, it is natural to first examine whether there exists an acyclic HP-completion set for a digraph $G$ of zero crossings, i.e., a *crossing-free acyclic HP-completion set* for $G$. In terms of an upward 2-page topological book embedding, this question is formulated as follows: given an embedded upward planar digraph $G$, determine whether there exists an upward 2-page book embedding of $G$ without spine crossings preserving $G$'s embedding.

In this paper we focus on crossing-free hamiltonian path completion sets for embedded upward planar digraphs. Our results include:

1. *Given an embedded st-digraph $G$ which has a crossing-free HP-completion set, we show that there always exists a crossing-free HP-completion set with at most two edges per face of $G$ (Theorem 1).*
   This result finds application to upward 2-page book embeddings. The problem of spine crossing minimization in an upward topological book embedding is defined with a scope to improve the visibility of such drawings. For the class of upward planar digraphs that always admit an upward 2-page book embedding (i.e. a topological book embedding without spine crossings) it make sense to define an additional criterion of visibility. When a graph is embedded in a book, its faces are split by the spine into several adjacent parts. It is clear that the visibility of a drawing improves if each face is split into as few parts as possible. This result implies that the upward planar digraphs which admit an upward 2-page book embedding also admit one such embedding where each face is divided to at most 3 parts by the spine.
2. *Given an embedded $N$-free upward planar digraph $G$, we show how to construct a crossing-free HP-completion set for $G$ (Theorem 3).* The class of embedded $N$-free upward planar digraphs is the class of embedded upward planar digraphs that does not contain as a subgraph the embedded $N$-graph of Figure 1.a. $N$-free upward planar digraphs have been studied in the context of partially ordered sets (posets) and lattices. The class of $N$-free upward planar digraphs contains the class of series-parallel digraphs which has been thorough studied in the context of book embeddings.
3. *Given a width-k embedded planar st-digraph $G$, we show how to determine whether $G$ admits a crossing-free HP-completion set (Theorem 5).* It follows that for fixed-width embedded planar $st$-digraphs, it can be tested in polynomial time whether there exists a crossing-free HP-completion set (and thus, a 2-page upward book embedding). The result is based on a reduction to the *minimum setup scheduling* problem.

This paper is organized as follows: In section 2 we present related previous work that has appeared in the literature. Section 3 presents the necessary terminology and notation while, Section 4 presents properties of crossing-free HP-completion sets for embedded $st$-digraphs that are used throughout the paper. The main results of the paper are presented in Sections 5, 6 and 7. In Section 5 we show that an embedded $st$-digraph which admits a crossing-free HP-completion set, also admits one with at most two edges per face. In Section 6 and 7 we examine $N$-free embedded upward digraphs and fixed-width embedded $st$-digraphs, respectively. We conclude in Section 8 with open problems.

## 2 Related Work

Motivated by parallel process scheduling problems, upward book embedding of acyclic digraphs and posets have been widely investigated [14,15,16,21]. If the book has only 2 pages, then the problem is called an *upward 2-page book embedding*. Alzohairi and Rival [2], showed that any upward planar series-parallel poset has an upward 2-page book embedding. In terms of graphs, it means that any upward planar series-parallel digraph without transitive edges has an upward 2-page book embedding. This result was improved by Giacomo et al. [9]. The authors provided a linear time algorithm, relaxing the constraint of non-transitive edges. On the same period an optimal algorithm that constructs an upward 2-page book embedding for $N$-free planar lattices was provided by Alzohairi [1]. Note that, due to Habib and Jerou [12], the class of $N$-free planar ordered sets can be considered as an extension of series-parallel ordered sets. Alzohairi [1] shows that any $N$-free planar lattice (or a planar $N$-free planar $st$-digraph without transitive edges) can be embedded in a 2-page book in $O(n)$ time. He also shows that any $N$-free planar ordered set that contains no covering four-cycle can be embedded in a 2-page book.

The problem of upward topological book embedding, i.e. upward book embeddings, where the edges are allowed to cross the spine, was introduced by Giordano et. al [10]. They showed that any planar embedded $st$-digraph has an upward topological book embedding where each edge crosses the spine at most once. The problem of spine crossing minimization was studied in terms of crossing minimization in acyclic hamiltonian path completion of an acyclic digraph (for short, Acyclic-HPCCM) [18,19]. The authors of [18,19] showed that the problem of spine crossing minimization in upward topological book embedding is equivalent to the problem of Acyclic-HPCCM for embedded planar $st$-digraphs. They also gave a linear time algorithm solving the Acyclic-HPCCM problem for the class of embedded outerplanar $st$-digraphs. The problem of Acyclic-HPCCM is relevant to the problem of *linear extension* (or, *jump number*) for posets. Both, the jump number and the Acyclic-HPCCM, ask to complete a given acyclic digraph to a hamiltonian acyclic digraph. The difference is that, in jump number the target is to minimize the number of edges added, while in Acyclic-HPCCM, to minimize the total number of crossings created by a newly added edges (provided that an upward planar embedding of a digraph is given). The problem of jump number has been shown to be NP-hard for some classes of orders(see [20],[22]). Up to our knowledge, its computational classification is still open for lattices. Nevertheless, polynomial time algorithms are known for several classes of ordered sets (see [4,5,6,7,24,25,11]).

Colbourn, Pulleyblank [6] and later Ceroi [3] studied the weighted version of jump number problem. In [6] this problem is referred as the problem of *minimum setup scheduling* (for short, MSS). The MSS is a problem of precedence constrained scheduling which appears to be different from other better studied problems of this field [23]. The authors of [6] give a polynomial time algorithm solving the MSS problem for the acyclic digraphs of bounded width. They observe also that the jump number problem can be presented as an instance of MSS. On the other, the MSS was shown to be NP-Complete for the class of two-dimensional orders [3].

## 3 Terminology and Notation

Let $G = (V, E)$ be a graph. Throughout the paper, we use the term *"graph"* we refer to both directed and undirected graphs. We use the term *"digraph"* when we want to restrict our attention to directed graphs. We assume familiarity with basic graph theory [13,8].

A *drawing* $\Gamma$ of graph $G$ maps every vertex $v$ of $G$ to a distinct point $p(v)$ on the plane and each edge $e = (u, v)$ of $G$ to a simple open curve joining $p(u)$ with $p(v)$. A drawing in which every edge $(u, v)$ is a a simple open curve monotonically increasing in the vertical direction is an *upward drawing*. A drawing $\Gamma$ of graph $G$ is *planar* if no two distinct edges intersect except at their end-vertices. Graph $G$ is called *planar* if it admits a planar drawing $\Gamma$.

An embedding of a planar graph $G$ is the equivalence class of planar drawings of $G$ that define the same set of faces or, equivalently, of face boundaries. A planar graph together with the description of a set of faces $F$ is called an *embedded planar graph*.

Let $G = (V, E)$ be an embedded planar graph, $E'$ be a superset of edges containing $E$, and $\Gamma(G')$ be a drawing of $G' = (V, E')$. When the deletion from $\Gamma(G')$ of the edges in $E' - E$ induces the embedded planar graph $G$, we say that $\Gamma(G')$ *preserves the embedded planar graph* $G$.

Let $G = (V, E)$ be a digraph. A vertex of $G$ with in-degree (resp. out-degree) equal to zero (0) is called a *source* (resp., *sink*). An *st-digraph* is an acyclic digraph with exactly one source and exactly one sink.

Traditionally, the source and the sink of an $st$-digraph are denoted by $s$ and $t$, respectively. An $st$-digraph which is planar and, in addition, embedded on the plane so that both of its source and sink appear on the boundary of its external face, is referred to as a *planar st-digraph*. In a planar $st$-digraph $G$ each face $f$ is bounded by two directed paths which have two common end-vertices. The common origin (resp., destination) of these paths is called the *source* (resp., *sink*) of $f$ and is denoted by $source(f)$ (resp., $sink(f)$). The leftmost (resp., rightmost) of these two paths is called a *left border* (resp., *right border*) of face $f$. The *bottom-left* (rest., *bottom-right*) edge of a face $f$ is the first edge on its left(resp., right) border. Similarly we define the *top-left* and the *top-right* edge of a face border. The *right(left) border* of an $st$-digraph is the rightmost(leftmost) path from its source $s$ to its sink $t$.

A new edge $e$ that is inserted to a face $f$ of a planar $st$-digraph $G$, with its origin and destination on the the left and right border of $f$, respectively, is called a *left-to-right oriented* edge. Analogously, we define a *right-to-left oriented* edge.

Following the terminology of posets, the digraph $G_N = (V_N, E_N)$, where $V_N = \{a, b, c, d\}$ and $E_N = \{(a, b), (c, b), (c, d)\}$ is called an *N-digraph*. Then, any digraph that does not contain $G_N$ as a subgraph is called an *N-free* digraph. This definition can be extended to embedded planar digraphs by insisting on a specific embedding. If we adopt the embedding of Figure 1.a. we refer to an *embedded N-digraph* while, if we adopt the embedding Figure 1.b. we refer to an *embedded И digraph*. An embedded planar digraph $G$ is then called $N$-free ($И$-free) if it does not contain any embedded $N$-digraph ($И$-digraph) as a subgraph. Figure 1.c shows an embedded $N$-free digraph. However, when its embedding is ignored, the digraph is not $N$-free since vertices $a$, $b$, $c$, $d$ comprise a $N$-digraph.

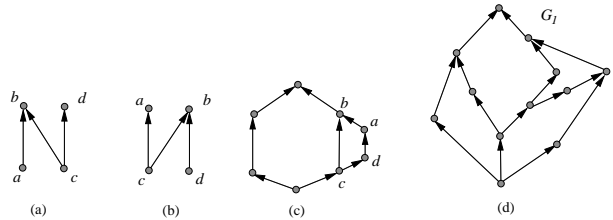

**Fig. 1:** (a) Embedded $N$-digraph. (b) Embedded $И$-digraph. (c) Planar digraph that is $N$-free if treated as an embedded planar digraph, but not $N$-free as a planar digraph. (d) An embedded $N$-free planar $st$-digraph $G_1$.

Let $G = (V, E)$ be an embedded planar $st$-digraph. The external face is split into two faces, $s^*$ and $t^*$. $s^*$ is the face to the left of the left border of $G$ while $t^*$ is the face to the right of the right border of $G$.

For each $e = (u, v) \in E$, we denote by $left(e)$ (resp. $right(e)$) the face to the left (resp. right) of edge $e$ as we move from $u$ to $v$. The *dual* of an $st$-digraph $G$, denoted by $G^*$, is a digraph such that: (i) there is a vertex in $G^*$ for each face of G; (ii) for every edge $e \neq (s, t)$ of $G$, there is an edge $e^* = (f, g)$ in $G^*$, where $f = left(e)$ and $g = right(e)$. If $G^*$ after this construction contains multiply edges, we substitute them by single edges. It is a well known fact that the dual graph $G^*$ of any planar $st$-digraph $G$, is also a planar $st$-digraph with source $s^*$ and sink $t^*$.

The following definitions were given in [10] for maximal planar $st$-digraph. Here we extend them for planar $st$-digraphs. Let $G = (V, E)$ be a planar $st$-digraph and $G^*$ be the dual digraph of $G$. Let $v_1^* = s^*, v_2^*, \ldots, v_m^* = t^*$ be the set of vertices of $G^*$ where the indices are given according to an $st$-numbering of $G^*$. By the definition of the dual $st$-digraph, a vertex $v_i^*$ of $G^*$ ($1 \leq i \leq m$) corresponds to a face of $G$. In the following we denote by $v_i^*$ both the vertex of the dual digraph $G^*$ and its corresponding face in digraph $G$. Face $v_k^*$ is called the *k-th face* of $G$. Let $V_k$ be the subset of the vertices of $G$ that belong to faces $v_1^*, v_2^*, \ldots, v_k^*$. The subgraph of $G$ induced by vertices in $V_k$ is called the *k-facial subgraph* of $G$ and is denoted by $G_k$.

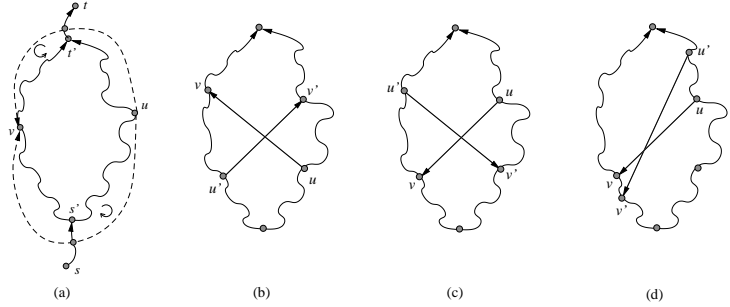

**Fig. 2:** (a) Construction for the proof of Property 1. (b)-(d) Construction for the proof of Property 3

The next lemma describes how, given an *st*-digraph $G$ and an *st*-numbering of its dual, $G$ can be incrementally constructed from its faces. The proof is identical to the proof given in [10] for maximal planar *st*-digraphs.

**Lemma 1.** *Assume a planar st-digraph $G$ and let $v_1^* = s^*, v_2^*, \ldots, v_m^* = t^*$ an st-numbering of its dual $G^*$. Consider the $k^{th}$-facial subgraph $G_k$ and the $k+1$-th face $v_{k+1}^*$ of $G$, ($1 \leq k < m$). Let $s_{k+1}$ be the source of $v_{k+1}^*$, $t_{k+1}$ be the sink of $v_{k+1}^*$, $s_{k+1}, u_1^l, u_2^l, \ldots, u_i^l, t_{k+1}$ be its left border, and $s_{k+1}, u_1^r, u_2^r, \ldots, u_j^r, t_{k+1}$ be its right border. Then:*

  a. *$G_k$ is a planar st-digraph.*
  b. *The vertices $s_{k+1}, u_1^l, u_2^l, \ldots, u_i^l, t_{k+1}$ are vertices of the right border of $G_k$.*
  c. *$G_{k+1}$ can be built from $G_k$ by an addition of a single directed path $s_{k+1}, u_1^r, u_2^r, \ldots, u_j^r, t_{k+1}$.* □

Let $G = (V, E)$ be an embedded planar *st*-digraph which has an acyclic crossing-free HP-completion set $S$. By $G_S = (V, E \bigcup S)$ we denote the HP-completed acyclic digraph and by $P_{G_S}$ the resulting hamiltonian path. Note that, as $S$ creates zero crossings with $G$, each edge of $S$ is drawn within a face of $G$ and, therefore, $G_S$ is a planar *st*-digraph.

## 4 Properties of Crossing-free Acyclic HP-completion Sets

In this section, we state some properties of crossing-free acyclic HP-completion sets that are useful in proving the main results of this paper.

**Property 1** *Assume an embedded planar st-digraph $G$ which has an acyclic crossing-free HP-completion set $S$. Let $f$ be a face of $G$ and $u$, $v$ be two vertices of $f$ that reside on $f$'s opposite borders. Then, any directed path from $u$ to $v$ contains at least one edge of $P_{G_S}$ that is drawn inside $f$ and is directed from the border $u$ resides on towards the opposite border (i.e., the border $v$ resides on).*

*Sketch of proof.* If we assume, for the sake of contradiction, that there is a path from $u$ to $v$ that does not pass through face $f$, then, we conclude that there is a cycle in the acyclic HP-completed graph $G_S$ (see Figure 2.a). The edge which is drawn inside $f$ belongs in the HP-completion set $S$ and, thus, in the resulting hamiltonian path $P_{G_S}$. □

**Property 2** *Assume an embedded planar st-digraph $G$ which has an acyclic crossing-free HP-completion set $S$. Let $u$ and $v$ be two vertices of $G$ which are connected by a directed path from $u$ to $v$. Then, the resulting hamiltonian path $P_{G_S}$ visits $u$ before $v$.*

*Proof.* If we assume that $P_{G_S}$ visits vertex $v$ before vertex $u$ then, we conclude that $G_S$ contains a cycle, a clear contradiction. □

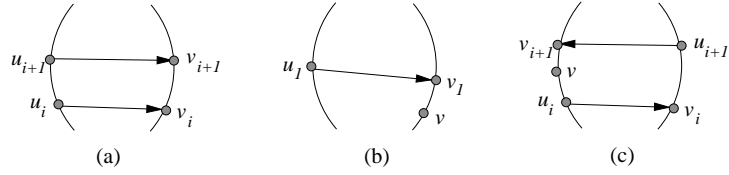

**Fig. 3:** (a)-(c) Constructions for the proof of Lemma 2

**Property 3** *Assume an embedded planar st-digraph $G$ which has an acyclic crossing-free HP-completion set $S$. Then, the edges of $S$ can all be drawn so that they do not cross each other.*

*Sketch of proof.* Suppose that there is an acyclic crossing-free HP-completion set $S$ which contains two edges $(u,v)$ and $(u',v')$ that cannot be drawn without crossing each other. Since $S$ is a crossing-free acyclic HP-completion set, then edges $(u,v)$ and $(u',v')$ are both drawn within the same face (see Figures 2.b-d). For the case shown in Figure 2.b, if we assume that $(u,v)$ is traversed before $(u',v')$, we infer that $v$ is traversed before $u'$, a contradiction due to Property 2. The remaining cases are treated similarly. □

**Lemma 2.** *Assume an embedded planar st-digraph $G$ which has an acyclic crossing-free HP-completion set $S$. Let $f$ be a face of $G$ and let $e_1, e_2, \ldots, e_k$ be the edges of $S$ which are drawn inside $f$, in the order they are traversed by the resulting hamiltonian path $P_{G_S}$. Then, it holds that:*
  a. *Edges $e_i$ and $e_{i+1}$ ($1 \leq i < k-1$) have opposite orientations.*
  b. *The destination of the $e_1$ is the lowermost vertex (other than $f$'s source) of either the left or the right border of $f$, while the origin of edge $e_k$ is the topmost vertex (other than $f$'s sink) of either the left or the right border of $f$.*
  c. *The origin of $e_i$ and the destination of $e_{i+1}$ ($1 \leq i < k-1$) are joined by a single edge of $G$.*

*Sketch of proof.*

(a) Note that this statement is meaningful only when $k > 1$, i.e., there exist at least two edges of $S$ that are drawn within $f$. Consider two consecutive edges, $e_i$ and $e_{i+1}$ in $f$ which are traversed in this order by the hamiltonian path $P_{G_S}$ and, for the sake of contradiction, assume that $e_i = (u_i, v_i)$ and $e_{i+1} = (u_{i+1}, v_{i+1})$ are both directed from the left to the right border of $f$ (see Figure 3.a). The hamiltonian path $P_{G_S}$ contains a directed path from $v_i$ to $u_{i+1}$. These vertices are placed in the opposite sides of $f$. So, by Property 1, the path from $v_i$ to $u_{i+1}$ contains at least one edge connecting the right border of $f$ with its left border. This edge can not be below $e_i$ or above $e_{i+1}$ as this leads to a contradiction due to Property 2. Thus, we have that there exists an edge in $S$ which is drawn within $f$, above $e_i$ and below $e_{i+1}$. This is a contradiction, since we assumed that $e_i$ and $e_{i+1}$ are consecutive edges of $S$ in face $f$.
(b) Without lost of generality, let $e_1 = (u_1, v_1)$ be left-to-right oriented (see Figure 3.b). For the sake of contradiction, we assume that the destination of the $e_1$ is not the lowermost vertex of the right border of $f$. Thus, there exists at least one vertex $v$, on the right side, which is placed below vertex $v_1$. By Property 2, vertex $v$ is visited before vertex $v_1$ and thus, $v$ is visited before $u_1$. So, the hamiltonian path $P_{G_S}$ contains a directed path from $v$ to $u_1$. Vertex $u_1$ is on the left side and thus, by Property 1, any path from $v$ to $u_1$ contains an edge in $f$ which is right-to-left oriented. Due to Property 2, this edge can not be above $e_1$ and thus, it must be below it. This is a clear contradiction, as we assumed $e_1$ to be the bottom edge of the HP-completion set drawn in $f$. The proof for $e_k$ is similar.
(c) Note that this statement is meaningful only when $k > 1$, i.e., there exist at least two edges of $S$ that are drawn within $f$. Let $e_i = (u_i, v_i)$ and $e_{i+1} = (u_{i+1}, v_{i+1})$ be two consecutive edges of $S$ which are drawn in $f$. Without lost of generality, we suppose that $e_i$ is left-to-right oriented (see Figure 3.c). Then, by statement (a) of this Lemma, we have that $e_{i+1}$ is right-to-left oriented. Assume, for the sake of contradiction, that $u_i$ and $v_{i+1}$, which are both on the left border of $f$, are not connected by an edge. Then, there exist a vertex $v$ on the left border of $f$ which is above $u_i$ and below $v_{i+1}$. Vertex $v$ has to be visited by $P_{G_S}$ after $u_i$ and before $v_{i+1}$ and, thus, after $v_i$ and before $u_{i+1}$. By Properties 1 and 2, we conclude that $e_i$ and $e_{i+1}$ are not consecutive edges of $S$ in $f$, a clear contradiction. □

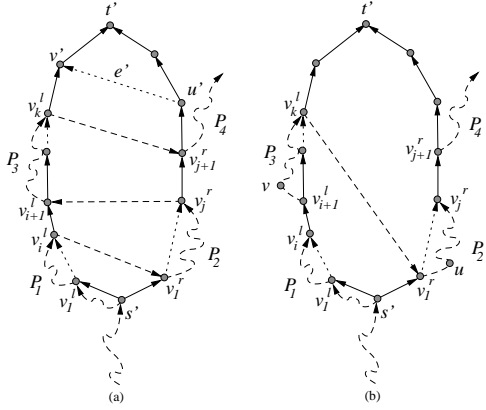

**Fig. 4:** (a) A crossing-free acyclic HP-completion set $S$ which places at least three edges to a face of an $st$-digraph. (b) An equivalent crossing-free acyclic HP-completion set $S'$ where the three edges $(v_i^l, v_1^r), (v_j^r, v_{i+1}^l), (v_k^l, v_{j+1}^r)$ were substituted by a single edge $(v_k^l, v_1^r)$.

## 5 Two edges per face are enough

In this Section, we prove that an embedded planar $st$-digraph $G$ which has a crossing-free acyclic HP-completion set, always admits a crossing-free HP-completion set with at most two edges per face of $G$. This result implies that the upward planar $st$-digraphs which admit an upward 2-page book embedding also admit one such embedding where each face is divided to at most 3 parts by the spine. This improves the quality of the book embedding drawing.

**Theorem 1.** *Assume an embedded planar st-digraph $G$ which has an acyclic crossing-free HP-completion set $S$. Then, there exists another acyclic crossing-free HP-completion set $S'$ for $G$ which contains at most two edges per face of $G$.*

*Proof.* Let $G_S$ be the acyclic HP-completed digraph and $P_{G_S}$ be the resulting hamiltonian path of $G_S$. We will show how to obtain from $S$ an acyclic zero-crossing HP-completion set $S'$ which has at most two edges per face of $G$.

Let $f = (s', v_1^l, \ldots, v_p^l, v_1^r, \ldots, v_m^r, t')$ be an arbitrary face of $G$ in which the acyclic HP-completion set $S$ places at least three edges. By Lemma 2, we infer that the first three edges of $S$ in $f$, in the order they are traversed by $P_{G_S}$, can be denoted by $S_f = \{(v_i^l, v_1^r), (v_j^r, v_{i+1}^l), (v_k^l, v_{j+1}^r)\}$, where, without lost of generality, we assumed that the first edge is left-to-right oriented. In Figure 4.a these three edges are denoted by dashed lines. Consider, for example, vertices $v_1^r$ and $v_j^r$ in Figure 4.a. The hamiltonian path $P_{G_S}$ can travel from $v_1^r$ to $v_j^r$ by either moving entirely on the right border of $f$, or, by visiting vertices at other faces of $G$. This second path that visits other faces besides face $f$ is drawn as a dashed curve from $v_1^r$ to $v_j^r$.

The three edges of $S_f$ split the hamiltonian path $P_{G_S}$ into 4 sub-paths[3], denoted by $P_1$, $P_2$, $P_3$ and $P_4$, where $P_1 = P_{G_S}[s \ldots v_i^l]$, $P_2 = P_{G_S}[v_1^r \ldots v_j^r]$, $P_3 = P_{G_S}[v_{i+1}^l \ldots v_k^l]$, and $P_4 = P_{G_S}[v_{j+1}^r \ldots t]$, and $s$ and $t$ are the source and the sink of $G$, respectively (see Figure 4.a).

We set $S' = S \setminus S_f \cup \{(v_k^l, v_1^r)\}$ (see Figure 4.b) and we will show that $S'$ is also an acyclic crossing-free HP-completion set for $G$, i.e., $G_{S'}$ is hamiltonian and acyclic, where $G_{S'}$ is a resulting graph when HP-completion set $S'$ is embedded in $G$.

By Lemma 2(c) we know that the edges $(v_i^l, v_{i+1}^l)$ and $(v_j^r, v_{j+1}^r)$ are present in $G$. Define path $P_{G_{S'}}$ to be formed by the concatenation of paths $P_1$, $P_3$, $P_2$ and $P_4$, that is, $P_{G_{S'}} = P_{G_S}[s \ldots v_i^l] \to$ $P_{G_S}[v^l_{i+1} \ldots v^l_k] \to P_{G_S}[v^r_1 \ldots v^r_j] \to P_{G_S}[v^r_{j+1} \ldots t]$. It is clear that path $P_{G_{S'}}$ is a hamiltonian path for $G_{S'}$.

---

[3] The notation $P[u \ldots v]$ indicates the subpath of $P$ starting at node $u$ and terminating at node $v$, or equivalently, the sequence of vertices in this subpath. We assume that the subpath is well defined, that is, $u$ appears before $v$ in $P$.

The only difference between $P_{G_{S'}}$ and $P_{G_S}$, is that in $P_{G_{S'}}$ the vertices of the sub-path $P_3$ are visited before the vertices of the path $P_2$. So a cycle could be created if there were two vertices $u$ and $v$, such that $u \in P_2$, $v \in P_3$ and there was a directed path from $u$ to $v$ in $G_{S'}$. However, if there is a directed path from $u$ to $v$ then there is a directed path from $v^r_1$ to $v^l_k$ (see Figure 4.b). By Property 1 we have that the path from $v^r_1$ to $v^l_k$ contains a right-to-left oriented edge $e' = (u', v')$ in $f$. The edge $(v^l_k, v^r_1)$ which we included to $S'$ is left-to-right oriented, and thus, $e'$ must be present in $S$ as well. By Property 3, $e'$ does not cross any edge of $S_f$. Also the edges of $S_f$ are the three lowest edges of $S$ in $f$, and thus, $e'$ is above $(v^l_k, v^r_{j+1})$ (see Figure 4.a). Now, it is easy to see that any path from $v^r_1$ to $v^l_k$ that passes through $e'$ creates a cycle in $G_S$. This is a clear contradiction as $S$ was supposed to be an acyclic HP-completion set.

So, we have substituted the three bottom edges of $S$ in $f$ by a single edge. By repeating this process, we can transform any odd number of edges to a single edge and any even number of edges to a pair of edges. □

## 6 Embedded $N$-Free Upward Planar Digraphs Always Have Crossing-Free Acyclic HP-completion Sets

In this Section, we study embedded $N$-free upward planar digraphs. We establish that any embedded $N$-free upward planar digraph $G$ has a crossing-free acyclic HP-completion set with at most one edge per face of $G$. Recall that the class of embedded $N$-free upward planar digraphs is the class of embedded upward planar digraphs that does not contain as a subgraph the embedded $N$-graph of Figure 1.a. For the class of $N$-free upward planar embedded digraphs, which is substantially larger than the class of $N$-free upward planar digraphs, we show that there is always a crossing-free acyclic HP-completion set that can be computed in linear time, thus improving the results given in [1,9]

**Theorem 2.** *Any embedded $N$-free planar st-digraph $G = (V, E)$ has an acyclic crossing-free HP-completion set $S$ which contains exactly one edge per face of $G$. Moreover, $S$ can be computed in $O(V)$ time.*

*Proof.* Let $G^*$ be the dual graph of $G$ and let $s^* = v^*_1, \ldots, v^*_m = t^*$ be the vertices of $G^*$ ordered according to an *st*-numbering of $G^*$. Let $G_{k-1}$ be the $(k-1)$-facial subgraph of $G$. By Lemma 1, $G_k$ can be constructed from $G_{k-1}$ by adding to the right border of $G_{k-1}$ the directed path forming the right border of $v^*_k$.

We prove the following stronger statement than the one in the theorem:

**Statement 1** *For any $G_k$ $(1 \leq k < m)$ there exists an acyclic crossing-free HP-completion set $S_k$ such that the following holds: Let $P_k$ be the resulting hamiltonian paht and let $e$ be an edge of the right border of $G_k$ that is also the bottom-left edge of a face $f \in \{v^*_{k+1}, \ldots v^*_m\}$. Then, edge $e$ is traversed by $P_k$.*

*Proof of Statement 1.* If $k = 1$, $G_1$ consist of a single path, that is the left border of $G$ (Figure 5.a). We let $S_1 = \{\emptyset\}$ and set $P_1$ to $G_1$. As all the edges of $G_1$ are traversed by $P_1$ it is clear that, any edge $e$ on the right border of $G_1$ that is also a bottom-left edge of any other face t is traversed by $P_1$.

Assume now that the statement is true for any $G_{k-1}$, $k < m$. We will show that it is true for $G_k$. Denote by $S_{k-1}$ a crossing-free acyclic HP-completion set of $G_{k-1}$ and by $P_{k-1}$ the produced hamiltonian path. Let $e$ be an edge on the right border of $G_{k-1}$ that is also the bottom-left edge of $v^*_k$. By the induction hypothesis, $P_{k-1}$ passes through $e = (s_k, v)$ (see Figure 5.b). Denote by $s_k$ and $t_k$ the source and the sink of $v^*_k$ respectively, and by $v^r_1, \ldots, v^r_{m_k}$ the vertices of the right border of $v^*_k$. By Lemma 1, $s_k$ and $t_k$ are vertices of the right border of $G_{k-1}$ and $G_k$ can be built from $G_{k-1}$ by adding the path $s_k, v^r_1, \ldots, v^r_{m_k}, t_k$ to it.

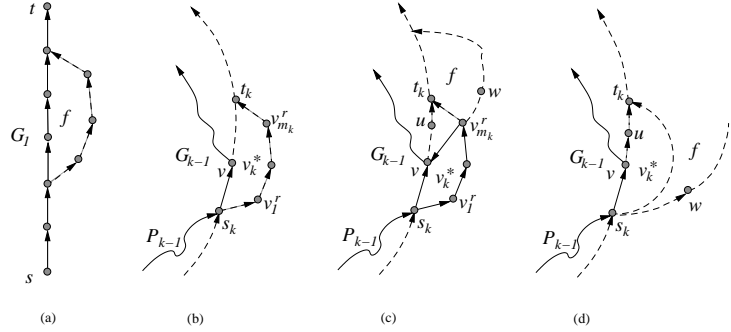

**Fig. 5:** (a) $G_1 = P_1$ and a face $f$. The bottom left edge of $f$ is traversed by $P_1$. (b) $G_{k-1}$ and $v_k^*$. $P_{k-1}$ is denoted by solid line. (c) A graph $G_k$ for the case that the right border of $v_k^*$ contains at least one vertex. The newly constructed $P_k$ is denoted by solid line. (d) A graph $G_k$ for the case that the right border of $v_k^*$ is a transitive edge.

Suppose first that $m_k \neq 0$ (i.e., the right border of $v_k^*$ contains at least one vertex). Set $S_k = S_{k-1} \bigcup \{(v_{m_k}^r, v)\}$, and $P_k = P_{k-1}[s \ldots s_k], v_1^r, \ldots, v_{m_k}^r, P_{k-1}[v \ldots t]$ (see Figure 5.c). It is clear that $P_k$ is a hamiltonian path of $G_k$. This is because $P_{k-1}$ is hamiltonian path of $G_{k-1}$ and $P_k$ traverses all newly added vertices. It is also easy to see that $S_k$ is acyclic: the edge $(v_{m_k}^r, v)$ which was added to $S_{k-1}$ creates a single directed path: from vertex $s_k$ to the vertex $v$, which were already connected by the directed edge $(s_k, v)$ in $G_{k-1}$.

We now show that the bottom-left edge $e$ of any $f \in \{v_{k+1}^* \ldots v_m^*\}$, where $e$ is also on the right border of $G_k$, is traversed by $P_k$. The only edge that was added to $G_{k-1}$ to create $G_k$ and is not traversed by $P_k$, is $e' = (v_{m_k}^r, t_k)$, that is, $e'$ is the last edge of the right border of $v_k^*$. If $e'$ is also the left bottom edge of a $f$ then the graph has an embedded $N$-digraph as a subgraph (see the subgraph induced by the vertices $u, t_k, v_{m_k}^r, w$ in Figure 5.c), a contradiction. Otherwise, if the bottom-left edge of $f$ coincides with any other edge of the right border of $v_k^*$, then the statement holds. If $f$ has its bottom-left edge on the right border of $G_{k-1}$ then, by the induction, a bottom left edge of $f$ is traversed by $P_{k-1}$ and, thus, by $P_k$.

Consider now the case where $m_k = 0$, that is, the right border of $v_k^*$ is a single, transitive edge (see Figure 5.d). In this case, no new vertex is added to $G_k$, so we set $S_{k+1} = S_k$ and $P_{k+1} = P_k$. Consider now a face $f \in \{v_{k+1}^* \ldots v_m^*\}$. If the bottom-left edge $e$ of $f$ is on the right border of $G_k$ and coincides with the transitive edge $(s_k, t_k)$, then $u, t_k, s_k, w$ form an embedded $N$-digraph (see Figure 5.d), a contradiction. So $e$ is not $(s_k, t_k)$ and, hence, it is an edge of the right border of $G_{k-1}$. So, by the induction hypothesis, $e$ is traversed by $P_{k-1}$ and hence by $P_k$. This completes the proof of the statement.

Having proved Statement 1, the theorem follows from the observation that $G_m = G$. The bound on the time needed to compute the crossing-free HP-completion set easily follows from the incremental nature of the described constructive proof. Figure 6 demonstrates the application of he algorithm implied by the constructive proof for the embedded $N$-free upward planar digraph of Figure 1.d. □

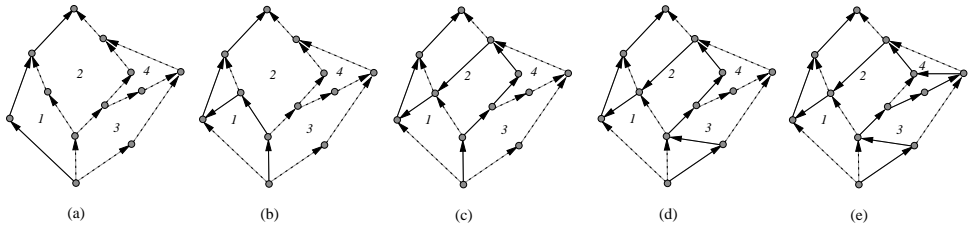

**Fig. 6:** (a)-(e) Construction of an acyclic zero-crossing HP-completion set and the corresponding hamiltonian path for the $N$-free planar $st$-digraph $G_1$ of Figure 1.d. The faces are numbered by an $st$-numbering of dual. Solid edges are the edges of hamiltonian path.

**Corollary 1** *Any И-free embedded planar st-digraph $G = (V, E)$ has an acyclic crossing-free HP-completion set $S$ which contains exactly one edge per face of $G$. Moreover, $S$ can be computed in $O(V)$ time.*

*Proof.* Just reverse the edges of $G^*$ and repeat the construction in the proof of Theorem 2.

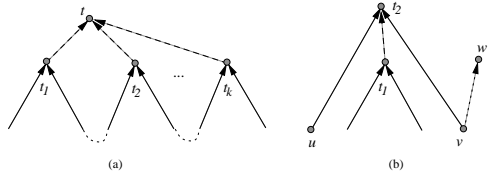

**Fig. 7:** (a)-(b) The construction for the proof of Theorem 3.

**Theorem 3.** *Any embedded N-free upward planar digraph $G = (V, E)$ has an acyclic crossing-free HP-completion set $S$. Moreover, $S$ can be computed in $O(V)$ time.*

*Proof.* We just prove that, any embedded N-free upward planar digraph $G$ can be transformed to an embedded N-free upward planar st-digraph $G'$ by the addition of few edges. Then, the result follows from Theorem 2.

Consider an upward planar embedding $\Gamma$ of $G = (V, E)$ that is N-free. If the outer face of $G$ contains more than one sink (source), then we add a new super-sink (super-source) vertex. Let $t_1, \ldots, t_k$ be the sinks of $G$ in the outer face. By adding a new vertex $t$ and by joining each $t_i$ to $t$ by an edge, the embedding $\Gamma$ of $G$ is preserved and remains N-free, because each $t_i$ $1 \leq i \leq k$ has out-degree zero (see Figure 7.a).

Let now some sink $t_1$ be placed in a inner face of $G$. Let $t_2$ be the sink of that face. We add edge $(t_1, t_2)$. The addition of the edge $(t_1, t_2)$ creates an embedded N-digraph only if there are edges $(v, t_2)$ and $(v, w)$ in $G$ with $(v, w)$ is the edge following $(v, t_2)$ (in counter clockwise order), out of $v$. But then, there is already an embedded N-digraph in $G$ (the digraph induced by the vertices $u, t_2, v, w$ in Figure 7.b). A clear contradiction, so $(t_1, t_2)$ can be added to $G$ without creating any embedded N-digraph as a subgraph. The sources are treated similarly. The transformation of $G$ into an st-digraph can be easily completed in linear time. □

## 7 Crossing-Free Acyclic HP-completion Sets for Fixed Width $st$-Digraphs

In this section we establish that for any embedded planar $st$-digraph $G$ of bounded width, there is a polynomial time algorithm determining whether there exists a crossing-free HP-completion set for $G$. In the case that such an HP-completion set exists, we can easily construct it.

A set $Q$ of vertices of $G$ is called *independent* if the graph incident to $Q$ has no edges. Following the terminology of partially ordered sets, we call *width* of $G$, and denote it by $width(G)$, the maximum integer $r$ such that $G$ has an independent set of cardinality $r$.

In Minimum Setup Scheduling (MSS) we are given a number of jobs that are to be executed in sequence by a single processor. There are constrains which require that certain jobs be completed before another may start; these constrains are given in the form of *precedence dag*. In addition, for each pair $i, j$ of jobs there is a *setup cost* representing the cost of performing job $j$ immediately after job $i$, denoted by $cost(i, j)$. The objective is to find a one-processor schedule for all jobs which satisfies all the precedence constrains and minimizes the total setup cost incurred.

The main idea of the result presented in this section is a simple application of an algorithm solving the *minimum setup scheduling* problem . Given a precedence dag $D$ and a matrix $C$ of costs, $s(D, C)$ denotes the total setup cost of a minimum cost schedule satisfying the constrains given by $D$. The next theorem follows from the complexity analysis given in [6].

**Theorem 4 ([6]).** *Given an n-vertex precedence dag $D$ of width $k$ and a matrix $C$ of setup costs, we can compute in $O(n^k k^2)$ time a setup cost $s(P, C)$ of minimum cost schedule, satisfying the constrains given by $P$.*

In the rest of this section, we show that given a planar $st$-digraph $G$ the problem of determining whether there is a crossing-free acyclic HP-completion set for $G$ can be presented as an instance of MSS.

Let $G = (V, E)$ is an embedded planar $st$-digraph. We define the setup cost matrix as follows. Set $C_G[i, j] = 0$ if $(v_i, v_j) \in E$ or $v_i$ and $v_j$ belong to the opposite borders of the same face of $G$, otherwise set $C_G[i, j] = 1$.

**Lemma 3.** *Let $G = (V, E)$ be an embedded planar st-digraph. Let also $s(G, C_G)$ be a setup cost of minimum cost schedule satisfying the constraints given by $G$ and setup costs given by $C_G$. $G$ has an acyclic crossing-free HP-completion set iff $s(G, C_G) = 0$.*

*Proof.* ($\Rightarrow$) Assume that $G$ has a crossing-free acyclic HP-completion set $S$ and the vertices in the sequence $v_1, v_2, \ldots, v_n$ are enumerated as they appear in the hamiltonian path which is created when $S$ is embedded on $G$. Then, the sequence $v_1, v_2, \ldots, v_n$ presents a schedule satisfying constraints given by $G$, otherwise an embedding of $S$ in $G$ would create a cycle. The setup cost for this schedule is $\sum_{i=1}^{n-1} C_G[i, i+1]$. We know that $S$ does not create any crossing with $G$. Therefore, any two successive vertices $v_i$ and $v_{i+1}$ of the resulting hamiltonian path are either connected by an edge of the graph or belong to the opposite borders of the same face, and thus, $C_G[i, i+1] = 0$. So, we have shown that there is a schedule of setup cost zero and, thus, $s(G, C_G) = 0$.

($\Leftarrow$) Assume now that $s(G, C_G) = 0$, i.e., there exists a one-processor schedule for the jobs represented by the vertices of $G$ which has total setup cost zero and satisfies the precedence constrains given by $G$. Let $v_1, v_2, \ldots, v_n$ be the jobs as they appear in this schedule. We construct the set of edges $S$ as follows: Consider any two successive jobs $v_i$ and $v_{i+1}$. If they are not connected by an edge $(v_i, v_{i+1})$ of $G$, then we add this edge to $S$. All the edges added to $S$ correspond to two jobs with setup cost zero, and hence represent edges which connect two vertices of the opposite borders of the same face. So we have that $S$ creates in $G$ a hamiltonian path which does not cross any edge of $G$.

We will show now that the addition of $S$ to $G$, does not create any cycles. Call $G'$ the graph resulting when the edges of $S$ are embedded on $G$. For the sake of contradiction, assume $G'$ contains a cycle. Then, there are two vertices $v_i$ and $v_j$ that are connected in $G'$ by two non-intersecting directed paths, from $v_i$ to $v_j$ and, from $v_j$ to $v_i$. Then the job corresponding to the vertex $v_i$ is executed by the schedule before the job corresponding to $v_j$ and vice versa. This is a contradiction. So $G'$ is acyclic.

We complete the proof with the observation that there can be no crossing among the edges of $S$ and, hence, the total number of crossing created by $S$ is zero. This fact follows directly from Property 3. □

**Theorem 5.** *Let $G$ be a planar st-digraph of width $k \in \mathbb{N}$. Then, in $O(k^2 n^k)$ time we can decide whether $G$ has an crossing-free HP-completion set. In the event that such a set exists, it can be easily computed in the same time bounds.*

## 8 Conclusions

We have studied the problem of crossing-free acyclic HP-completion for planar embedded $st$-digraphs and for upward planar embedded digraphs. We have showed that if a planar embedded $st$-digraph has a crossing-free HP-completion set then it also has a crossing-free HP-completion set with at most two edges per face of a given drawing. An upward 2-page book embedding implied by such HP-completion set has an improved visibility as each face is split into at most three parts. For the class of $N$-free upward planar embedded digraphs, which is substantially larger than the class of $N$-free upward planar digraphs, we show that there is always a crossing-free acyclic HP-completion set that can be computed in linear time, thus improving the results given in [1,9]. Finally, we proved that there is a strong relation between the jump number problem and the problem of existence of an crossing-free acyclic HP-completion set, i.e. both can be treated as an instance of the weighted jump number problem.

The list of open problems includes:

1. Finding a full characterization of upward planar embedded digraphs that have an acyclic crossing-free HP-completion set. This problem seams to be easier than the problem of characterizing acyclic digraphs or posets with pagenumber 2 [17,21].
2. Studying the computational complexity of the Acyclic-HPCCM problem. Is the Acyclic-HPCCM problem more difficult than the problem of examining whether there is an acyclic crossing-free HP-completion set?